\documentclass[conference,a4paper]{IEEEtran}
\usepackage{cite}
\usepackage{amsmath,amssymb,amsfonts}
\usepackage{tabularx,booktabs}
\usepackage{algorithmic}
\usepackage{algorithm}
\usepackage{graphicx}
\usepackage{textcomp}
\usepackage{xcolor}
\def\BibTeX{{\rm B\kern-.05em{\sc i\kern-.025em b}\kern-.08em
		T\kern-.1667em\lower.7ex\hbox{E}\kern-.125emX}}
\begin{document}
	
	\def\tablename{TABLE}
	
	    \title{Power Adaptation for Suborbital Downlink with Stochastic Satellites Interference}
 
	    \author{
        \IEEEauthorblockN{
        Yihao He, 
        Juntao Ma, 
        Zhendong Peng, 
        and Gang Wu} \\
		\IEEEauthorblockA{University of Electronic Science and Technology of China, Chengdu, China}  \\
		Email: wugang99@uestc.edu.cn
        }

        %\author{
        %\IEEEauthorblockN{
        %Yihao He\IEEEauthorrefmark{1}, 
        %Juntao Ma\IEEEauthorrefmark{1}, 
        %Zhendong Peng\IEEEauthorrefmark{2}, 
        %and Gang Wu\IEEEauthorrefmark{1}} \\
		%\IEEEauthorblockA{\IEEEauthorrefmark{1}University of Electronic %Science and Technology of China, Chengdu, China}  \\
        %\IEEEauthorblockA{\IEEEauthorrefmark{2}University of British %Columbia, Vancouver, Canada}  \\
		%Email: yhhe@std.uestc.edu.cn, wugang99@uestc.edu.cn
        %}

	\maketitle

	\begin{abstract}
 
        This paper investigates downlink power adaptation for the suborbital node in suborbital-ground communication systems, which are subject to extremely high reliability and ultra-low latency communications requirements. 
        The problem is formulated as a power threshold-minimization problem, where interference from satellites is modeled as an accumulation of stochastic point processes on different orbit planes, and hybrid beamforming (HBF) is considered. 
        To ensure Quality of Service (QoS) constraints, the finite blocklength regime is adopted. 
        Numerical results show that the required transmit power of the suborbital node decreases as the elevation angle at the receiving station increases. 
        
	\end{abstract}

	\section{Introduction}
	
	    Space-earth communications is one of the important scenarios in 6G mobile networks \cite{2021_6G}.
	    Currently, commercial suborbital missions using reusable rockets and suborbital spacecrafts are becoming increasingly frequent, which makes the study of suborbital communications meaningful. 
	    Suborbital vehicles can be modeled as nodes in sky-earth networks.
	    Due to its extremely high speed and frequent maneuvers, the transmission of real-time data between the suborbital node (SN) and the ground station needs to meet a stringent Quality of Service (QoS). 
	    It's lucid that Whether QoS can be guaranteed or not is directly related to the signal-to-noise ratio (SNR) \cite{2018_shechangyang}. 
        However, the SN can be located in a large altitude range, and the channel characteristics are similar to those of low-orbit satellites, adjacent or even the same frequency is often used between satellites and the SN. 
        Also, with the establishment of satellite constellations, interference from satellites may influence downlink transmission of the SN more frequently, which is worth considering. 
        To overcome interference and ensure the downlink QoS, a direct way is to increase the transmission power of the SN to enhance the signal-to-interference plus noise ratio (SINR) of the ground station. 
        Despite this, blindly increasing the transmit power of the SN may lead to security problems of eavesdropping on the ground, increasing of hardware cost, and other potential defects. 
        Furthermore, it is challenging to describe these large amounts of cumulative interference definitely as the characteristics of satellites are difficult to capture with various observation locations. 
	    There is only a little literature on power adaption in suborbital-ground transmission. 
	    Still, according to the scenario of a suborbital downlink transmission, the QoS requirements of the SN can be measured by extremely high reliability and ultra-low latency (ERLLC). 
	    Specifically, the overall packet error probability is $10^{-9}$, and the packet size is $20$ bytes \cite{2021_6G}. 
 
	    For ERLLC, the authors of \cite{2018_shechangyang} introduce a compact power adaptation strategy under QoS requirements. 
	    They optimized the packet dropping and power adaptation policy to minimize the transmit power under QoS constraints, where a cross-layer power adaptation strategy is conducted. 
	    As the packet length is small, the classical Shannon channel capacity becomes unsuitable. 
	    Therefore, the effective bandwidth is introduced and the finite blocklength coding theorem is adopted to model and optimize the channel capacity according to \cite{2010finite_blocklength,2018_finitelengthcoding}. 
        When the SN also adopts short packet communications, this modeling approach is very meaningful for suborbital communication scenarios.
	    Additionally, the authors of \cite{2022_TWC_Joint_Power_MCarrier} propose a joint power adaptation strategy for multi-carrier duplex and consider a method to maximize reliability via finding the optimal SINR. 
	    Moreover, \cite{2022_TWC_Joint_Resource_Edge_Computing} investigated the power adaptation with edge computing with short packet transmission, and the locally optimal solution can greatly reduce the computation complexity. 
	
	    However, when it comes to suborbital-ground transmission, various limitations make the above-mentioned methods inapplicable, and the reasons can be summarized as follows. 
	    First, as the multi-antenna array is deployed on both sides of the SN and the receiver, beamforming in space needs to be considered to enhance power efficiency. 
	    Second, the discard strategy to deal with the error packets in \cite{2018_shechangyang} can be simplified accordingly.	
	    Third but most importantly, interference from satellites in outer space needs to be considered, more precisely, in a stochastic way. 
	
	    In our work, we propose a power adaptation strategy for the SN's downlink transmission under suborbital-ground systems with stochastic satellite interference. 
	    Consistently, We adopt the finite blocklength regime to ensure the QoS constraints motivated by \cite{2018_shechangyang}. 
	    Specifically, we propose a transmission-insisting short-packet ALOHA (TISPA) policy to simplify the process at the receiver.
	    In the space domain, an effective Hybrid beamforming (HBF) design from \cite{2020_TVT_HBF} is adopted at satellites and the receiving station to eliminate side-lobe interference and lower the cost as much as possible. 
	    Thanks to this, the fluctuation of the main beam is smaller, and the side lobes are better suppressed. 
	    On the other hand, due to the poor diffraction ability at Ku-band, the channel of satellite interference is considered as line of sight (LoS) \cite{2021_GC_satellites}.
        Therefore, we assume the interference in the main lobe of the beam is the main culprit affecting the reception. 
	    Additionally, a stochastic model is constructed to describe the interference on different orbit planes and directions by the stochastic point process according to \cite{2012_SGbookMartinHaenggi,2021_SGinterference}. 
	    Finally, the optimization problem is derived and solved, followed by simulation results. 
	
	\section{System Model}
\begin{figure}[t]
		\begin{centering}
			\includegraphics[width=1\columnwidth]{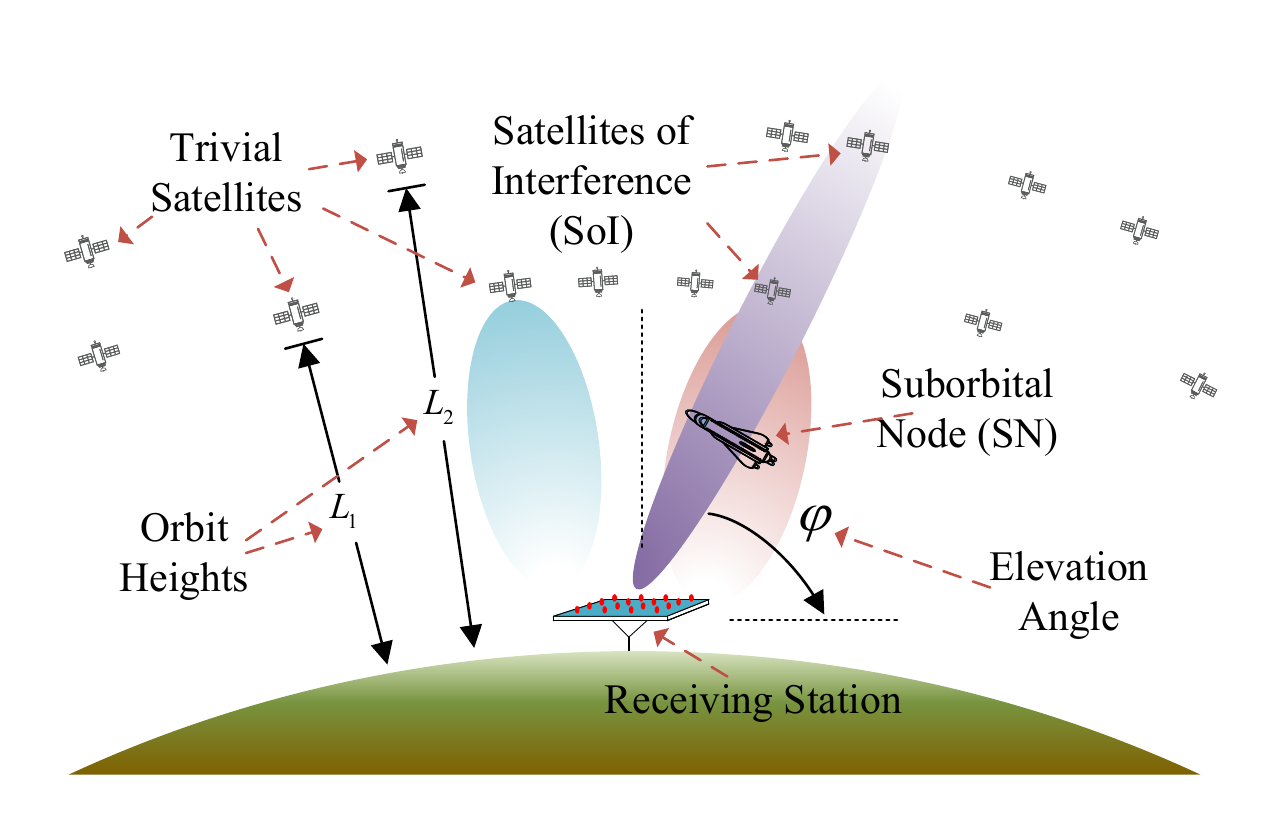}
			\par\end{centering}
		\centering{}\caption{\label{fig:Total distribution in URLLC}Downlink of the SN with satellites interference.}
	\end{figure}

	    As shown in Fig. \ref{fig:Total distribution in URLLC}, we consider the downlink of the SN with a receiving station at the ground. 
	    We assume that the SN generates commands and flight parameters into short packets of Poisson Process (PP) queuing to be transmitted through the downlink. 
        Same as ERLLC devices on the ground, the SN will process signaling data in real-time. 
        Due to the high frequency and comparably narrow band, we assume that the downlink channel is quasi-static and frequency non-selective within a frame inspired by \cite{2014_NtinShortpkt,2021_GC_satellites}. 
	    The downlink communication is under QoS requirements, i.e., packet error probability $\varepsilon^{c}$, queue violation probability $\varepsilon^{q}$, and maximum tolerable delay $D_{max}$ \cite{2018_shechangyang}. 
     
	    Besides, there are randomly distributed satellites with different orbital heights ${L_i}$ in the background of the SN and within the main lobe of the beam from the receiving station, i.e., satellites of interference (SoI). 
	    From the view of SoI, the power of their main lobes affect the SINR at the receiving station. 
	    The elevation angle of the beam at the receiving station is denoted by ${\varphi}$. 
	
	\subsection{Power Adaptation for the SN}

	    The downlink effective bandwidth can be guaranteed by increasing the transmit power of the SN \cite{2010finite_blocklength}. 
        Still, it is limited by a power upper limit due to the hardware restriction, etc. 
        Hence the transmit power strategy of the base station is segmented, i.e.
	\begin{equation}
		P=\begin{cases}
			P_{t}(g) & g\geq g_{th}\\
			P_{u} & g\leq g_{th}
		\end{cases}, 
        \label{eq:Pt_case}
	\end{equation}
	    where $P_{t}$ denotes the minimum transmitted power needed to meet the QoS, and $P_u$ means the maximum transmitted power allowed by the transmitter in the SN. 
        $g_{th}$ is the equivalent threshold of the small-scale fading coefficient $g$ (i.e. when $g$ is larger than $g_{th}$, $P_{t}$ can be smaller than $P_{u}$, and in this case the downlink transmission is carried out normally). 
        
        $P_u$ is relatively large to meet most of the conditions, but there are still rare occasions caused by interference that require higher $P_{t}$. 
        Accordingly, there is a probability that $g$ is smaller than $g_{th}$ denoted as $\varepsilon^t$. 
	    Here suppose the SN still sends the downlink packet with maximum transmitted power $P_{u}$ if $\varepsilon^{c}$ cannot be guaranteed, which is dubbed TISPA. 
        Thus the essential approach to reduce costs and optimize the global power strategy is to obtain the minimum $P_{u}$.
        
	    After the derivation in the next section, the optimization for TISPA is formulated as
	\begin{equation}
		\begin{aligned} & \min_{\varepsilon^{c},\varepsilon^{q},\varepsilon^{t}}P_{u}=\frac{N_{0}B+\mathbb{E}(I)}{g_{th} G}\nu\\
			& \begin{array}{r@{\quad}r@{}l@{\quad}l}
				{\rm {s.t.}C_{1}:} & \varepsilon=\varepsilon^{c}+\varepsilon^{q}+\varepsilon^{t}\leq\varepsilon^{QoS}\\
				C_{2}: & \varepsilon^{c},\varepsilon^{q},\varepsilon^{t}\geq0,\\
			\end{array}
		\end{aligned}
		\label{eq:PuProblem}
	\end{equation}
	    where we mark
	\begin{equation}
		\nu={{\rm exp}(\frac{(E_{B}\eta{{\rm ln}2)}}{B}+\sqrt{\frac{1}{T_{f}B}}Q^{-1}(\varepsilon^{c}))-1\label{eq:niu_{t}emp}}
	\end{equation}
        for simplicity. 
        $\eta$ is the packet size in bits per packet. 
        $N_{0}$ represents the unilateral power spectral density (PSD) of Gaussian white noise, $B$ is the bandwidth of the channel used, and $Q^{-1}$ means the inverse of the Gaussian Q-function.
	    Additionally, the path loss, the period of a frame, and the power of stochastic interference caused by satellites are denoted by $G$, $T_{f}$, and $I$ respectively. 
        $\mathbb{E}(I)$ stands for the expectation of total stochastic interference from satellites. 
	    Explicitly, (\ref{eq:PuProblem}) is the core problem of the power adaptation at an SN. 
	
	\subsection{Stochastic Interference from Satellites}

	    The interference of each satellite is jointly determined by the channel gain and spatial gain. 
	    The commonly used Ku-band channel of SoI is considered as LoS \cite{2021_GC_satellites}. 
	    Assuming SoI transmitting full-time, $I$ is a cumulative power from satellites distributions of different orbits, i.e. 
	\begin{equation}
		I = P_{\rm{SN}}\sum\limits_i {\sum\limits_{x \in {\Phi _i}} {{f_{\rm{PL}}}(x){G_{tx}}(x){G_{rx}}(x)} } , \label{eq:I}
	\end{equation}
	    where $i$ is the index of orbits, $\Phi={x_1,x_2,...}\subset\mathbb{R}^{2}$ represents the stochastic point process of interference satellites on the orbital plane. 
        We assume that the launch power of each satellite is subject to uniform distribution within a certain range.
        By a simple derivation of the compound PP, the expectation of launch power $P_{\rm{SN}}$ is what we only need. 
        Additionally, the path loss from each satellite, the power gain of the antennas at satellites, and the power gain of the antennas at the receiving station are denoted by $f_{\rm{PL}}$, $G_{tx}$, and $G_{rx}$ respectively. 

	    $I$ in (\ref{eq:I}) is a function of the stochastic point position ${x}$. 
	    The distribution on the orbital plane is modeled as a homogeneous Point Poisson Process (PPP) in a spherical surface $\mathbb{R}^{2}$ when observed at the geocentric angle or the length of the satellite orbit.

	\section{Derivation and Algorithm}
	
	\subsection{Coding Rate and Effective Bandwidth of the Downlink}

        One theorem used in this part is the maximum coding rate of short packets \cite{2010finite_blocklength}, which is an indicator to measure the downlink of the SN. 
        With approximation, it can be written as
	\begin{equation}
		{R^*}(n,{\varepsilon ^c}) \approx \frac{B}{{\ln 2}}[\ln (1 + \gamma ) - \sqrt {\frac{\sigma }{n}} {Q^{( - 1)}}({\varepsilon ^c})],
        \label{eq:R*origin}
	\end{equation}
	    where $\gamma$ stands for the SINR at the receiving station. 
	    $\sigma$ is the approximated channel dispersion 
	\begin{equation}
		\sigma  = 1 - {(1 + \gamma )^{ - 2}}
	\end{equation}
        according to \cite{2010finite_blocklength}. 
        $\sigma\approx1$ when $\gamma$ is high. 
	    Thus, for a fixed $\sigma$, the optimal coding rate $R^*$ will increase as block-length $n$ tends to be a large value. 
	    With $n=T_fB$, and $\eta$ denote the packet size, $R^*$ can be expressed as
        \begin{equation}
		R^{*}(\varepsilon^{c})\approx\frac{T_{f}B}{\eta{{\rm ln}2}}{{\rm ln}(1+\frac{g G P}{N_{0}B+I})-\sqrt{\frac{1}{T_{f}B}}Q^{-1}(\varepsilon^{c})}.
        \label{eq:R*simplify}
	\end{equation}

        Besides, the minimum service rate of the SN should be larger than its effective bandwidth determined by the queuing process of downlink packets. 
	    To establish the relationship between SINR and the transmission rate of short packets at the SN, it is necessary to adopt the effective bandwidth $E_B(\theta)\triangleq\Delta(\theta)/\theta$. 
	    And can be expressed as $E_{B}(\theta)=\Omega/2\theta T_{f}$ by extending the similar scenario in \cite{2018_shechangyang} to PP, where
	\begin{equation}
		\Omega=\frac{-2T_f{\rm{ln}}(\epsilon^q)}{D_{max}}, \label{Omega}
	\end{equation}
	    in which $D_{max}$ represents the maximum tolerable delay. 
        $\theta$ denotes the rate of the exponential tail of the queue length distribution, which is a variable in the function of minimum envelope rate \cite{1994_queue}. 
        And then $E_{B}(\theta)$ can be derived as
	\begin{equation}
		\begin{cases}
		    E_{B}(\theta)=\frac{{\rm ln}(1/\varepsilon^{u})}{D_{max}{{\rm ln}(\frac{T_{f}{\rm ln}(1/\varepsilon^{u})}{\lambda_q D_{max}}+1)}}\\
			\theta={\rm ln}(\frac{T_{f}{{\rm ln}(1/\varepsilon^{u})}}{\lambda_q D_{max}}+1)
		\end{cases},\label{eq:EB&theta}
	\end{equation}
	    where $\varepsilon^{u}$ denotes the upper bound of delay violation probability. 
	    $\lambda_q=N_u*T_f$ is the density of data generated, and $N_u$ represents the packet arrival rate. 

        \subsection{Optimal Strategy Deduction}

	    According to the effective bandwidth, with $E_B T_f \le R^*$, we consider the lower bound as the service rate and convert (\ref{eq:R*simplify}) into a function of $P$, i.e.
	\begin{equation}
		{P_t} \buildrel \Delta \over = \frac{{{N_0}B+I}}{{gG}}\nu,
        \label{eq:Pt}
	\end{equation}
	and $P\ge{P_t}$. 
        Moreover, $\nu$ is the same in (\ref{eq:niu_{t}emp}), which is actually the SINR at the receiver when the SN is transmitting. 
	    Consistent with (\ref{eq:Pt_case}), ${P_t}$ is the minimum required to transmit power to meet the QoS of the downlink. 
	    In this way, the SN transmits packets with a rate of effective bandwidth. 
        Since $R^{*}$ depends on the maximum transmission power $P_{u}$ when $g$ is fixed to $g_{th}$, the maximum achievable rate of $R^{*}$ can be denoted as
	\begin{equation}
		R_{th}=\frac{T_{f}B}{\eta{{\rm ln}2}}{{\rm ln}(1+\frac{g_{th} G P_{u}}{N_{0}B+I})-\sqrt{\frac{1}{T_{f}B}}Q^{-1}(\varepsilon^{c})}.
	\end{equation}
	
	    Since the downlink channel from the SN is different from other satellites and has a lower altitude, the small-scale fading $g$ needs to be considered differently from the satellite-ground transmission.
	    Thanks to \cite{2014_NtinShortpkt}, the finite blocklength theorem is available in the quasi-static multi-antenna channel within a frame of the downlink transmission. 
        The probability distribution function (pdf) of $g$ is a Nakagami distribution \cite{2020_TCOM_UAV}, which is 
	\begin{equation}
		f_{g}(x)=\frac{1}{(N_{t}-1)!}x^{N_{t}-1}e^{-x},\label{eq:fg_x}
	\end{equation}
	where $N_{t}$ denotes the number of transmitting antennas. 
	    The TISPA scheme of (\ref{eq:Pt_case}) considers the probability of $g<g_{th}$ occurring is small if there is a relatively large $P_{u}$. 
	    Consistent with (\ref{eq:PuProblem}), the probability of $g\leq g_{th}$ is written as
	\begin{equation}
		\varepsilon^{t}=\int_{0}^{g_{th}}f(g)dg,\label{eq:epsilon_t}
	\end{equation}
	    which can be considered as the error probability introduced by TISPA.
	    According to the expression of $g_{th}$ from (\ref{eq:Pt}) and (\ref{eq:epsilon_t}), the total transmission error probability $\varepsilon$ can be approximated as 
	\begin{equation}
		\begin{split}
			\varepsilon  = 1 - (1 - {\varepsilon ^q})(1 - {\varepsilon ^c})(1 - {\varepsilon ^t})\\
			\approx {\varepsilon ^q} + {\varepsilon ^c} + {\varepsilon ^t}.
		\end{split}
	\end{equation}
	
	    Basically, minimizing $P_{u}$ is equivalent to minimizing $\nu$ under fixed $\varepsilon^{t}$ according to (\ref{eq:PuProblem}), i.e. 
	\begin{equation}
		\begin{aligned} & \min_{\varepsilon^{c},\varepsilon^{q}}\frac{{{\rm ln}(1/\varepsilon^{q})\eta{\rm ln}2}}{D_{max}{{\rm ln}(\frac{T_{f}{{\rm ln}(1/\varepsilon^{q})}}{\lambda_q D_{max}}+1)B}}+\sqrt{\frac{1}{T_{f}B}}Q^{-1}(\varepsilon^{c})\end{aligned}
		.\label{eq:subproblem}
	\end{equation}
	    By solving the second-order differential of the expression above, it can be proved that the sub-problem (\ref{eq:subproblem}) is convex w.r.t. $\varepsilon^{c}$ and $\varepsilon^{q}$. 
	
	    After the specific analysis above, the solution to the optimization problem in (\ref{eq:PuProblem}) can be summarized in Algorithm 1.
	    \begin{algorithm}[t]{}\label{alg1}
		\begin{centering}
			\caption{Finding $P_u$ from (\ref{eq:subproblem}) under TISPA}
			\par\end{centering}
		\begin{algorithmic}[]		
			\REQUIRE  $N_u$, $T_f$, iterate times $J$, and QoS limitations;	\\	
			\textbf{Phase I:} Pre-calculation of $G$ by definition and $\lambda_q$. 
			\begin{itemize}
				\STATE Obtain path gain $G$ and $\lambda_q=N_uT_f$; 
			\end{itemize}		
			\textbf{Phase II:} Solve the sub-problem (\ref{eq:subproblem}). 		
			\begin{itemize}
				\STATE Initialize $\epsilon_q$ with 5 pre-set values;			
				\FOR{$j$ from 0 to $J$}			
				\STATE Get $\epsilon_c$ from each $\epsilon_q$, and calculate (\ref{Omega});			
				\STATE Set the bound of $\theta$ from (\ref{eq:EB&theta}); 			
				\STATE Find $\theta$ by bisection method, and calculate $E_B$ from (\ref{eq:EB&theta});			
				\STATE Calculate the exponent in (\ref{eq:niu_{t}emp});			
				\STATE Find $\epsilon_q$ by bisection method, then obtain $\epsilon_c$ according to QoS, and get a corresponding $g_{th}$ from (\ref{eq:PuProblem}), where $\epsilon_t$ is obtained from (\ref{eq:epsilon_t}); 			
				\STATE Renew the bound of $P_u$ in this loop;			
				\STATE Find the optimal $P_u$ by bisection method; 			
				\ENDFOR 
			\end{itemize}		
			\ENSURE $P_u$		
		\end{algorithmic}
	\end{algorithm}
	
	\subsection{Calculation of Stochastic Interference}
\begin{figure}[t]
		\begin{centering}
			\includegraphics[width=0.7\columnwidth]{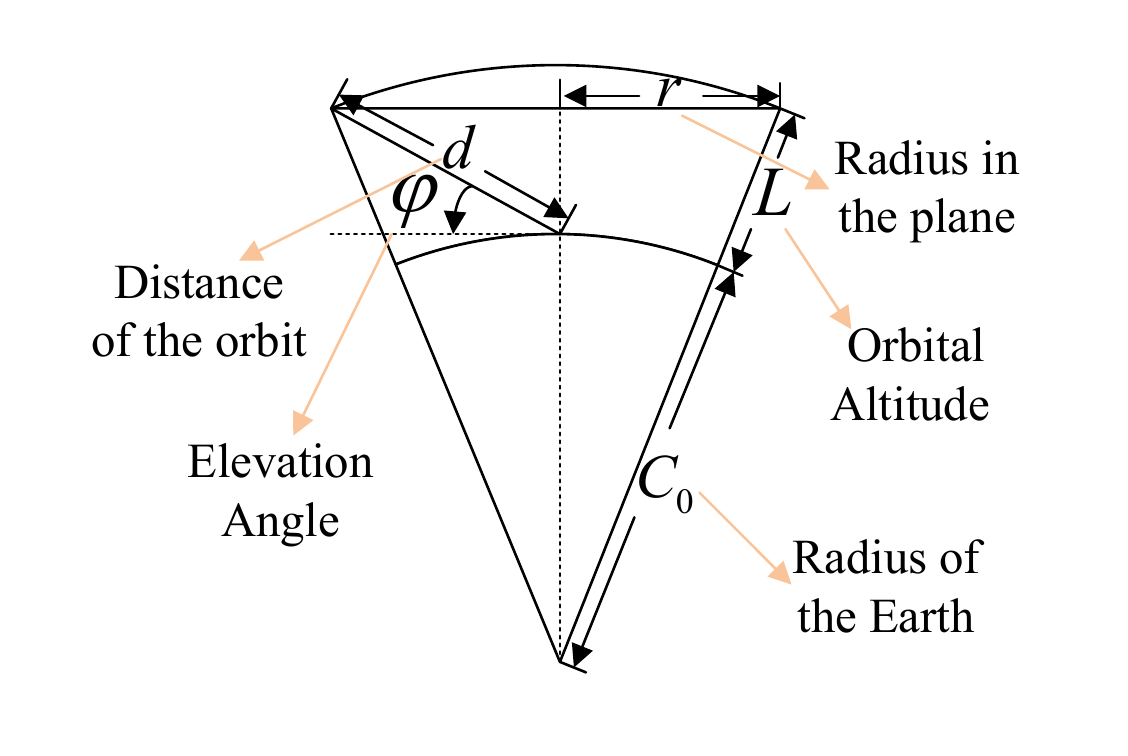}
			\par\end{centering}
		\centering{}\caption{\label{fig:system_orbit}Geometric relationship considered.}
	\end{figure}

	    The distribution of satellites is homogeneous w.r.t. the geocentric angle.
        Thus, the density distribution of interference is related to the geometric relationship of the observation points, i.e., azimuth angle $\vartheta$ and elevation angle $\varphi$ from the receiving station. 
	    $L$ represents the height of the satellite orbital plane, the radius of the earth is $C_0$, and $d$ denotes the transmission distance to a certain spatial position on the orbital plane. 
	    The geometric relationship under each observation domain is shown in Fig. \ref{fig:system_orbit}.
	    Since the area where the main lobes of SoI and the receiving station overlap are small, it can be approximately considered that the corresponding elevation angle of SoI and the receiving station that generates interference share the same $\varphi$.
	
	    The stochastic point process $\Phi=\{ x_1,x_2,...\}$ is a space on the orbital plane with two dimensions, and the distribution of satellites can be converted to the union process $\Phi_a=\{\vartheta_1,\vartheta_2,...\}$ and $\tilde\Phi =\{\varphi_1,\varphi_2,...\}$ w.r.t. the azimuth and elevation angle respectively. 
        For an equivalent array antenna covering the hemisphere, the main lobe of the beam is considered to be fixed in the middle of the azimuth. 
	    Assuming that the base of the receiving station can rotate horizontally while tracking the SN. 
        Hence, the distribution of the path loss changes in azimuth does not cause a difference in distance, i.e., $f_{\rm{PL}}$ is only a function of $\varphi$. 
        Then the array antenna degenerates into an equivalent linear array. 
	    Thus, the antenna gain at the receiving station is fixed for different azimuth angles, i.e., both $G_{tx}$ and $G_{rx}$ are only functions of $\varphi$. 
	    Based on the most widely used low earth orbit (LEO), medium earth orbit (MEO), and geosynchronous earth orbit (GEO), we consider the geometric distribution of satellites as orbital planes at 3 altitudes to generalize satellites in adjacent regions at that altitude. 
	    For more accurate results, the division of the orbital plane can be more detailed, which will not be discussed here. 
	Thereupon, (\ref{eq:I}) can be specifically written as
	\begin{equation}
		I=P_{\rm{SN}}\sum\limits_{i = 1}^3 {\sum\limits_{\varphi \in {\tilde\Phi_i}} {{f_{\rm{PL}}}(\varphi){G_{tx}}(\varphi){G_{rx}}(\varphi)} },\label{eq:I_phi}
	\end{equation}
	    where $i$ is the index of the 3 orbits mentioned. 
	
	    As the range of antenna scanning is small when interference occurs, it is approximately considered that the stochastic distribution on the satellite orbits under the $\varphi$ observation domain is quasi-homogeneous. 
	    The propagation in the Ku-band is LoS, and the path loss in free space is inversely proportional to the square of $d$, i.e., $f_{PL}\propto (1/d)^2$. 
        However, there is still a part of the energy in the atmosphere, and the equivalent path loss on the whole link is slightly higher. 
	    Therefore, the path loss of the interfering signal is considered a segmented function of the track height, which is organized as a function of transmission distance according to the geometric relationship, i.e.,
	\begin{equation}
		{f_{\rm{PL}}}(d) = {(\frac{{L_{r}d}}{L})^{ - \alpha }} + {(\frac{{{d^2} - L_{r}d}}{L})^{ - {\alpha _0}}},\label{eq:fPL}
	\end{equation}
	    where both path loss factors ${\alpha_0}$ and ${\alpha}$ in outer space and inside the atmosphere are considered, and $L_{r}$ denotes the reference height of the boundary of the atmosphere.
	
	    In the satellite-ground link, besides high energy efficiency, a low enough fluctuation is needed, i.e., not affecting the gain of stochastic interference in the space domain. 
	    An HBF array design with better side lobe suppression and smaller main lobe ripple is adopted. 
        The design process needs to solve an optimization problem with passband flattening and hybrid decomposition with details in \cite{2020_TVT_HBF}. 
	    Since the beam of common satellites is fixed or moves within a small range, we assume that their main beams are fixed in the center. 
	    A certain scanning elevation angle is selected for the receiving station, and then the corresponding gain can be determined by the HBF. 
	
	    Subsequently, the function of the stochastic point process in (\ref{eq:I_phi}) is converted into a function w.r.t. $d$, and the space of $\varphi$ is also mapped to the space of $d$, i.e., 
	\begin{equation}
		I=P_{\rm{SN}}\sum\limits_{i = 1}^3 {\sum\limits_{d \in {\Phi^*_i}} {{f_{\rm{PL}}}(d){G_{tx}}(d){G_{rx}}(d)} },\label{eq:I_d}
	\end{equation}
	    where $\Phi^*=\{d_1, d_2,...\}$ is an inhomogeneous point process in the distance domain as $d$ is unevenly distributed.
	    Still, on a relatively small area of the orbital plane, the distribution is homogeneous when viewed from a distance $r$ from the center of the orbital plane (i.e. the radius in the orbital sub-plane, which is introduced as a reference variable for a homogeneous distribution w.r.t. subset $X \in \mathbb{R}$ of $\Phi_r =\{r_1, r_2,...\}$ and is equivalent to the expression in (\ref{eq:I_phi})). 
	    Let $Y \in \mathbb{R}$ be a subset of $\Phi^*$, and according to the geometric relationship, the mapping function from the original subset of $\Phi_r$ is $f(x) = \sqrt {\parallel x{\parallel ^2} + {L^2}} $.
	
	    With known geometric relations, the measure of this distribution in the space w.r.t. $d$ is
	\begin{equation}
		{\Lambda ^*}(Y) = {\lambda _d}\pi {r^2} = {\lambda _d}\pi ({d^2} - {L^2}),
	\end{equation}
	    where $\lambda _d$ denotes the density of distribution in the orbit that can be calculated from the number of satellites $n_L$ in orbit and $L$. 
	    And then, the density measure is ${\lambda ^*}(r) = 2{\lambda_d}\pi r = 2{\lambda_d}\pi \sqrt {{d^2} - {L^2}}$.
	
	    Finally, the measure function under the original distribution can be expressed as
	\begin{equation}
		\Lambda (\partial x) = 2\pi {\lambda _d}\sqrt {{d^2} - {L^2}} \partial d,\label{eq:measure_d}
	\end{equation}
	    where the partial differential representation is used to prevent confusion of $d$.
	
	    According to Campbell's Theorem \cite{2012_SGbookMartinHaenggi}, the expectation of a function of a random point process can be expressed as the R-S integral w.r.t. the density measure.
	    Thus, combining (\ref{eq:I_d}) and (\ref{eq:measure_d}), the expectation of $I$ is obtained, i.e. 
	\begin{equation}
		\mathbb{E}(I) = 2\pi {\lambda _d}P_{\rm{SN}}\sum\limits_{i = 1}^3 {\int_{{d_{li}}}^{{d_{ui}}} {F(d)\sqrt {{d^2} - {L^2}} \partial d} }, \label{eq:E_I}
	\end{equation}
	where $F(d) = {{f_{{\text{PL}}}}(d){G_t}(d){G_r}(d)}$.
	    The upper limit $d_{ui}$ and lower limit $d_{li}$ of the integral is determined by the actual overlap of the beams on the i-th orbit. 
	    Due to the expression of $f_{\rm{PL}}$ and the numerical relationship of $G_r$ and $G_r$, the integral can be solved numerically.
	
	    Employ a uniform interval of $\cos \varphi $ to get a vector of ${\sqrt {{d^2} - {L^2}} }/d$, and then perform linear interpolation on the expression to obtain the numerical relationship of $G_r$ and $G_t$ with $d$. 
	    Eventually, the integral can be solved by the complex trapezoidal integral with no more discussion here.

	\section{Numerical Results}
	
        In this section, we present numerical results that demonstrate the effectiveness of our proposed power adaptation scheme for the SN in the presence of interference from satellites. 
        We provide details of the simulation results and list the parameter settings used in Table \ref{table:parameters}.
     
	\begin{table}[t]
		\centering
		\caption{Parameters in calculation \cite{2018_shechangyang,2014_NtinShortpkt,2021_6G}}
		\begin{tabular}{lcc}
			\toprule
			Parameters & Symbol & Value  \\
			\midrule
			Frame length  & $T_{f}$ & $0.1$ ms \\
			Packet arrival rate  & $N_u$ & $10000$ pkt/s \\
			Transmission error probability & $\varepsilon$ & $10^{-9}$ \\
			Maximum waiting time & $D_{max}$ & $0.1$ ms \\
			Path loss factor in the atmosphere& $\alpha$ & $2.5$ \\
			Path loss factor in space& $\alpha_0$ & $2$ \\
			Packet length & $\eta$ & $160$ bits \\
			Bandwidth & $B$ & $10$ MHz \\
			Frequency & $f_c$ & $12$ GHz \\
			Power spectral density & $N_{0}$ & $-150$ dBm \\
			Path Loss & $G$ & $ -20{\rm{lg}}\frac{\lambda_w}{4\pi d_0}-10\alpha{\rm{lg}}\frac{d_k}{d_0}$ dB \\
			Transmit power of a satellite & $P_{\rm{SN}}$ & 10 W \\
			Satellites orbit altitude & $L$ & $[350,1000,4000]$ km \\
			Satellites in orbit & $n_L$ & $[60000, 8000, 600]$ \\
			\bottomrule
		\end{tabular}
		\label{table:parameters}
	\end{table}
	
	\begin{figure}[t]
		\begin{centering}
			\includegraphics[width=1\columnwidth]{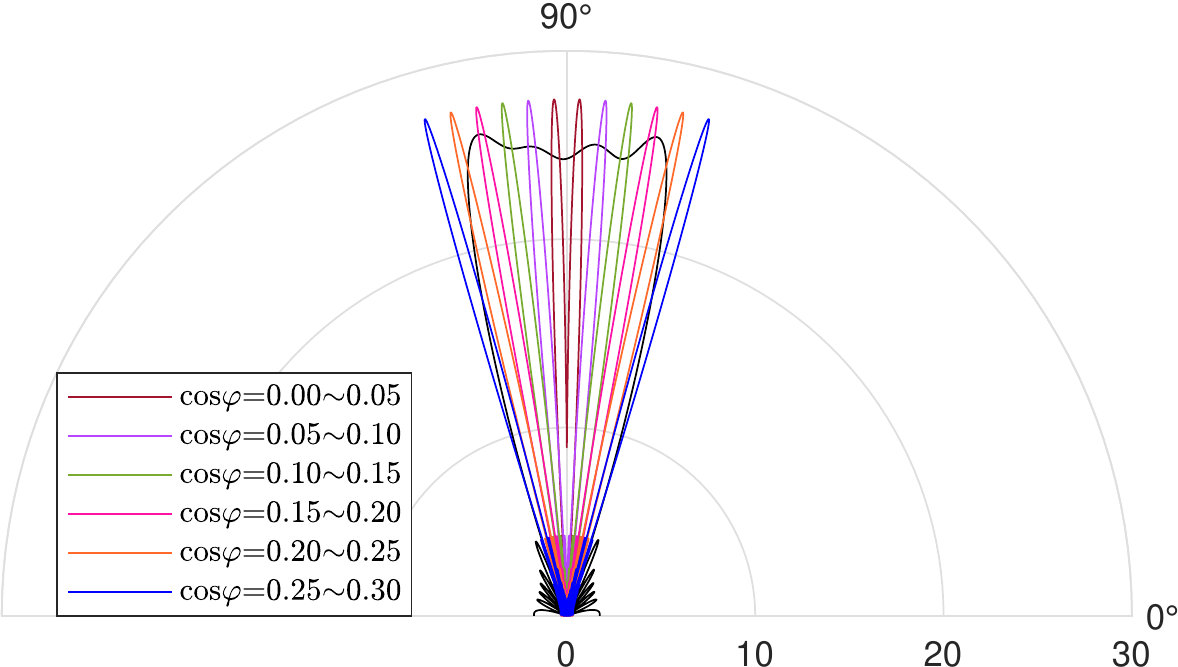}
			\par\end{centering}
		\centering{}\caption{\label{fig:beam_pattern}Equivalent overlapping gain of the beam with LEO satellites.}
	\end{figure}

        In Fig. \ref{fig:beam_pattern}, we show the overlaps of the SN beam with the satellites on the LEO. 
        The width of the satellite beam is fixed to $\arccos (\left| {0.28} \right|)$. 
        Overlap is only calculated when the beams are present on both sides \cite{2020_TVT_HBF}. 
        Additionally, there is a small ripple inside the main lobe of the satellite, which influences the total gain. 
        Similar effects are observed on the MEO and the GEO.
	
	\begin{figure}[t]
		\begin{centering}
			\includegraphics[width=1\columnwidth]{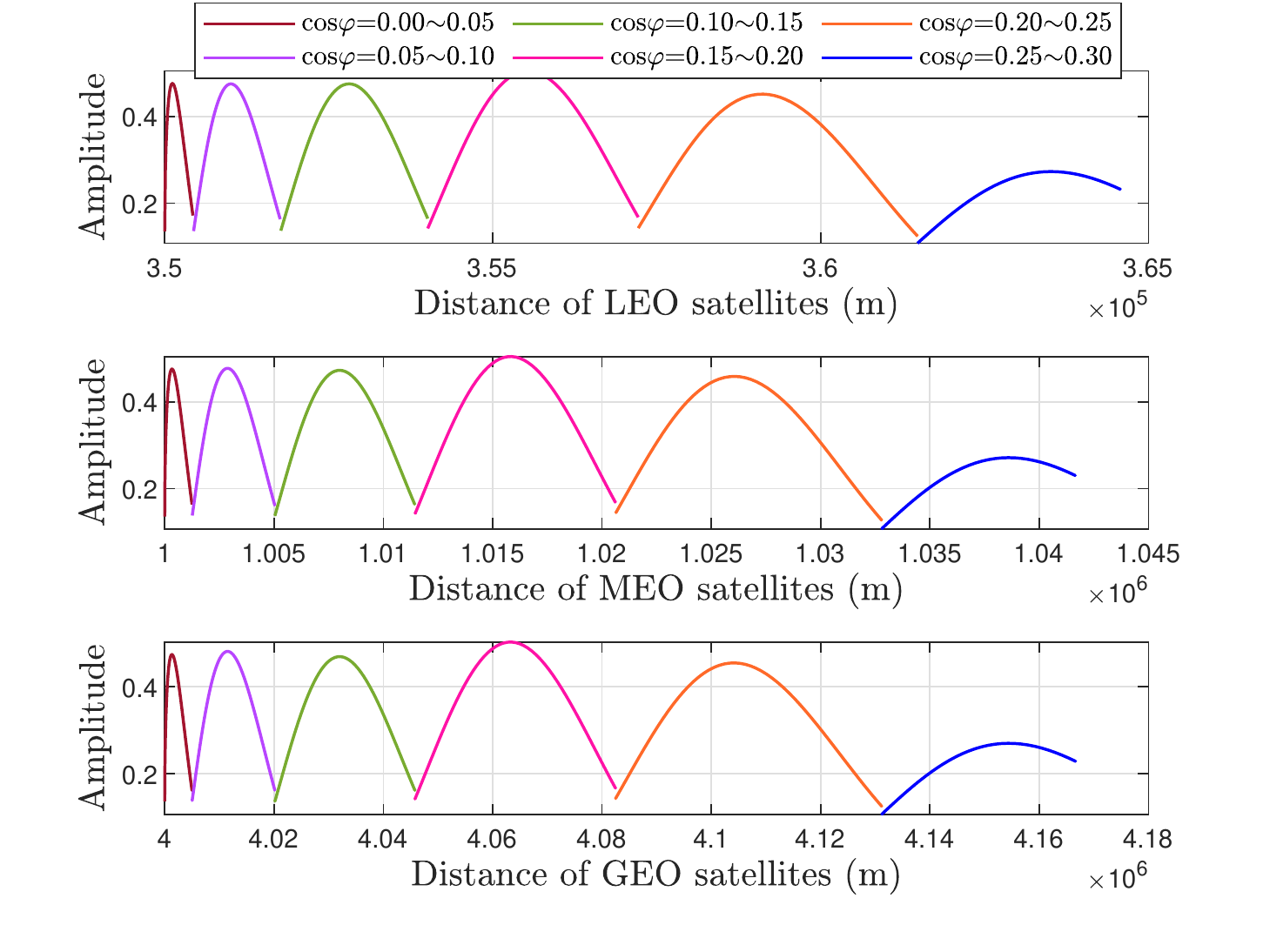}
			\par\end{centering}
		\caption{\label{fig:Go_Amp}Amplitude of overlap Gain by distances of different orbits.}
	\end{figure}

        We display the overlap gains at various distances away from the receiving station in Fig. \ref{fig:Go_Amp}. 
        The gains of the two beams are calculated using samples that are uniform in $\varphi$. 
        To solve the integral in (\ref{eq:E_I}), we transform $G_r$ and $G_t$ into the relationship of equidistant sampling w.r.t. $d$. Receiving at those three orbits with different main lobe angles, we scan the relationship between the overlapped gain of the transceivers and the distance. 
        The results show that as the elevation angle $\varphi$ decreases, the overall gain peaks when the cosine of the elevation angle is approximately between 0.15 and 0.20. Moreover, the decline of $\varphi$ indicates that the SN is moving away from directly above, and a larger zone of space is scanned with more satellites. 
        As the receiving beam of the receiving station moves out of the beam of the satellite, the interference gain drops sharply and almost disappears.
	
	\begin{figure}[t]
		\begin{centering}
			\includegraphics[width=1\columnwidth]{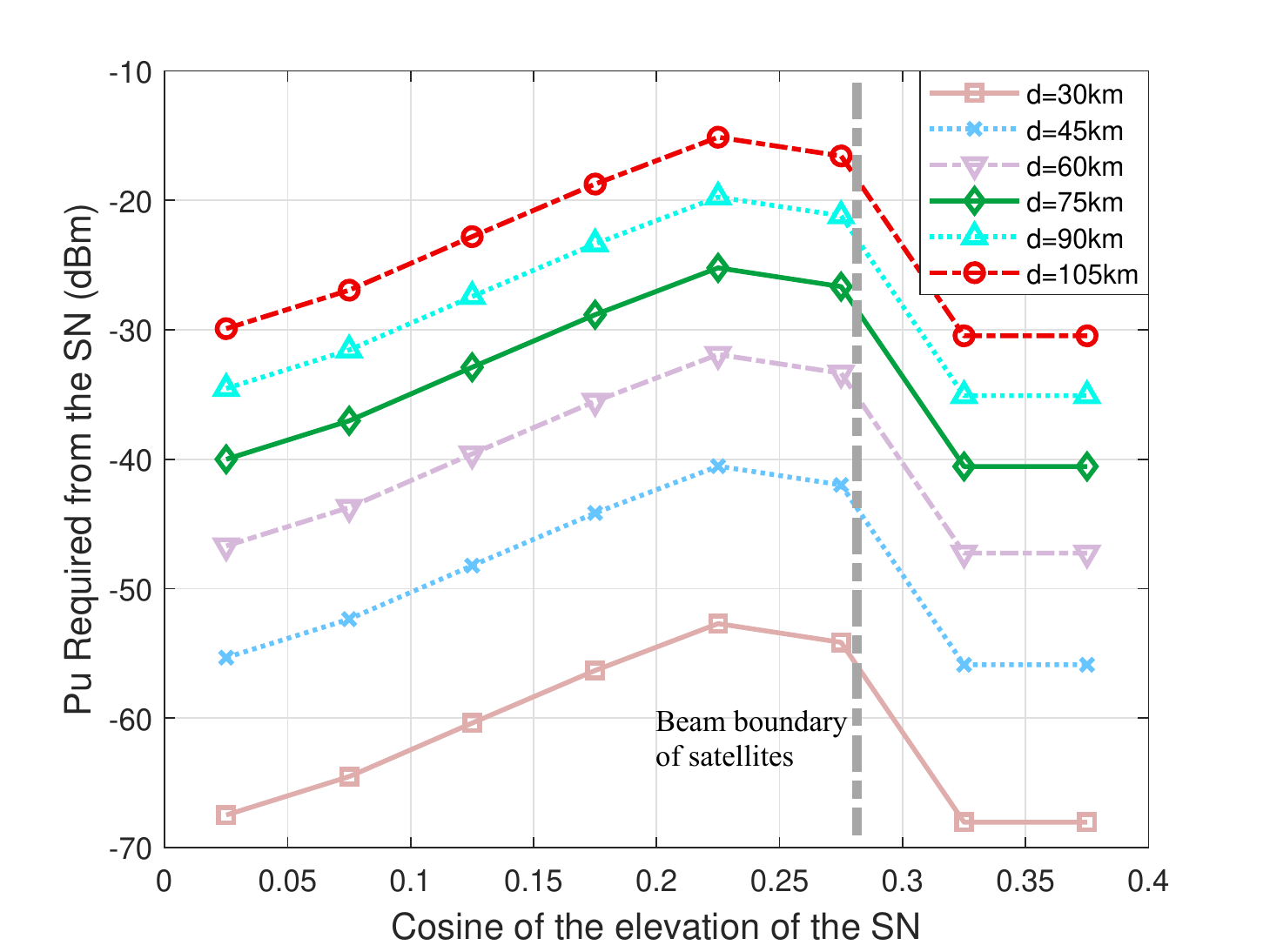}
			\par\end{centering}
		\caption{\label{fig:simu}Performance comparison with different conditions.}
	\end{figure}

        In Fig. \ref{fig:simu}, we present the power threshold $P_u$ required at the SN to maintain QoS under interference with diverse beam elevation angles scanned from the receiving station. 
        The results are obtained using our proposed optimization model. 
        It is evident that as the SN's height increases, the transmission distance increases and the power requirement for the downlink connection increases accordingly. 
        Simultaneously, as the elevation angle deviates from directly above, the distance of the SN link also increases, and the beam at the receiving station introduces a larger interference distribution area, leading to increased interference.
        It is clear that the closer the satellite is to the center of the main lobe of the receiving station, the greater the overall gain is. 
        Their relationship scales with the orbital height. 
        Moreover, as the received beam moves away from the satellite beam direction, the interference drops sharply to zero. 
        Finally, we note that the scanning of the beam requires the accumulation of two main lobes, which is similar to convolution in the space domain.

	\section{Conclusion}

        Based on the research conducted in this paper, we have proposed a TISPA policy to simplify the process for the downlink of the SN in the presence of interference from satellites. 
        The finite blocklength regime was used to satisfy QoS constraints, and the interference from satellites was modeled by a stochastic point process, taking into account the HBF. 
        We formulated an optimization problem and solved it to determine the optimal power adaptation strategy for the downlink of the SN. The numerical results demonstrate that when the SN is within the directional range of the beam of SoI, the power threshold needs to be increased as the deviation from directly above and the height of the SN increase. 
        Future research improvements may include using updated information-theoretic models to improve the flow of optimization problems, considering the integration of sensing and communication of SN and satellite signals, and using more reasonable sub-orbital channel modeling in the modeling of satellite interference, etc.

	\bibliographystyle{IEEEtran}
	% \bibliography{main} % 和tex同名
 

\end{document}